\documentstyle{article}
\def\be{\begin{equation}}
\def\ee{\end{equation}}
\def\bea{\begin{eqnarray}}
\def\eea{\end{eqnarray}}
\def\la{\label}
\def\ci{\cite}
\def\lr{\left}
\def\rr{\right}

\begin{document}

\begin{flushright}
OUTP-93-34P \\
\end{flushright}
\vspace{15mm}

 \begin{center}
 {\large\bf Phenomenological implications from  moduli fields in strings}
\\
 \vspace{7mm}
{\large  A. de la Macorra}\footnote{From 1 January 1994 E-mail:
macorra@teorica0.ifisicacu.unam.mx}\\ [5mm]
{\em Department of Physics, University of Oxford,\\
  1 Keble Rd, Oxford OX1 3NP}\\ [8mm]
 \end{center}
\vspace{2mm}
\begin{abstract}
\noindent

We study  some  phenomenological consequences of  having moduli fields with
large
vacuum expectation values (v.e.v.). The v.e.vs of the moduli  are dynamically
determined
  once supersymmetry is broken in 4D string theories.
The study constraints the possible
Yukawa interactions and modular weights for matter fields.

\end{abstract}
\renewcommand{\thefootnote}{\arabic{footnote})}
\setcounter{footnote}{0}

\newpage

	String theory provides with the only possibility, up till now, of unifying all
interactions \ci{r1}.
As formulated in the critical dimensions it has only one free parameter which
is
taken as the Planck mass. But once it is compactified to four dimensions much
of its uniqueness is lost because there are a great number of consistent string
vacua \ci{r25}-\ci{r36}. Nevertheless it is possible to study general 4D string
models by concentrating
on  properties shared by all models and consider the model dependent
quantities as free parameters.

One of the generic features of 4D string vacua is the existence
of dilaton and moduli fields.
The dilaton field gives the gauge coupling constant at the string
scale while the moduli parameterizes the geometry and
complex structure of the compactified dimensions.
All kinetic terms of the chiral matter fields are not canonical
but have  a $\sigma$-like structure and are moduli dependent.

Another generic property is the invariance of the effective
Lagrangian under duality symmetry \ci{r7}. Although it has only been
proved to be exact to all orders in perturbation theory \ci{r38} one
expects that non-perturbative effects will respect duality. This
symmetry is intimately related to the contribution of the
infinite number of Kaluza-Klein modes, always present in string
theory. These modes become relevant in the low energy effective
theory for small values of
the compactified radius given by the real part of the (1,1)
moduli.  In the simplest case, the moduli fields transform
under duality as an element of the SL(2,Z) group \ci{r5,r34} and in this
paper we will only consider those moduli.
If the standard model derives from 4D string theory its
parameters  are  moduli dependent and
their value is determined once their functional dependence  and
vacuum expectation value (v.e.v.) of these fields is known.
Duality symmetry has been very useful in constraining the form
of this  effective Lagrangian and determining the gauge coupling
constant, Yukawa couplings  and the  possible  string vacua.
In particular, the Yukawa couplings have been explicitly calculated for
orbifold compactifications  and some Calabi-Yau manifolds \ci{r60}.  In any
concrete example the expression of the Yukawa couplings can be
determined but in practice we still don't know how to get the standard model
 from the 4D string models. It is therefore useful to work in a model
independent
way  to study and reduce the number of  phenomenological viable string vacua.
 The Yukawa couplings  are given, in general, as  modular functions of the
moduli fields.   Since the Yukawa couplings are moduli dependent  it is
possible to have a fifth force from the interaction between the moduli and
ordinary quarks and leptons. This interaction could be experimentally detected
only if
the mass of the exchanged particle is small enough to render a long range
interaction. The mass of the the real part of the moduli $T_{i}$ is of order of
the gravitino mass while the imaginary part of  $T_{i}$ is exponentially
suppressed with respect to the gravitino mass. The exponent is proportional to
the v.e.v. of the real part of $T_{i}$ and therefore if  it has a large  v.e.v.
a fifth force could be observable.

The values of the moduli and dilaton v.e.v. are only fixed after
supersymmetry is broken. One expects supersymmetry to be broken by a gaugino
condensate \ci{r9} in the hidden sector and transmitted to the visible sector
(where the usual quarks and leptons live) via gravity.  At tree level the
v.e.v. of the moduli is of order one \ci{r49,r50} and in such a case
the mass of the imaginary part of $T_{i}$ is comparable to the gravitino mass
which is expected to be of the order of  the electroweak scale. On the other
hand, once loop corrections
of  the strong binding effects, which lead to  gaugino condensation, are taken
into account supersymmetry is broken with a single gaugino condensate   and
the v.e.v. of some moduli is
much larger \ci{r52,r56}. This result is also welcome because it renders an
unification scale, which is moduli dependent, of the order of $10^{16}\, GeV$
and  consistent with the unification  of the standard model  gauge coupling
constants \ci{r54,r59}.

Before discussing the Yukawa  interactions
we present the low energy 4D string model.   In order to study the breaking of
SUSY
one determines the effective potential and   its  vacuum.
The effective interaction involving the
gaugino bilinear can be obtained
by demanding the complete Lagrangian
 to be anomaly free under the R-symmetry  under which the
gauginos transform non-trivially.
The low energy degrees of freedom are then
the dilaton field $S$, moduli fields $T_{i}$, chiral
matter fields $\varphi_{i}$, gauge fields
and the Goldstone
mode $\Phi$ associated with the spontaneously broken   R-symmetry plus
their supersymmetric partners. In orbifold compactification there are
always three diagonal moduli
whose real parts represent the size of the
compactified complex plane and here we will only consider these moduli.
The 4D string model is given by an N=1 supergravity theory and it is specified
once the Kahler potential $G=K+ln\frac{1}{4}|W|^{2}$ and the gauge kinetic
function $f$ are given.
The Kahler potential, superpotential and gauge kinetic functions are
\ci{r37}-\ci{r51}
\be
K=-ln\lr(S+\bar{S}+2\Sigma_{i}(k_{0}\delta_{GS}^{i}-b'^{i}_{0})lnT_{ri}\rr)-\Sigma_{i}ln(T_{ri})+K^{i}_{i}|\varphi_{i}|^{2},
\la{d38}\ee
\be
W_{0}=\Pi_{i}\eta^{-2}(T_{i})\,\,\Phi\,+\,W_{m}
\la{d39}\ee
and \ci{r52}
\be
f=f_{0}+\frac{2}{3}b_{0}ln(\Phi)
\la{d40}\ee
respectively,  where $T_{r}\equiv T+\bar{T}$ and  $k_{0}$ is the  level of the
corresponding Kac-Moody algebra. $W_{m}$ is the superpotential for the chiral
matter
superfields
and the gauge kinetic function at the string scale is
given by \ci{r17}-\ci{r18},\ci{r39}
\be
f_{0}=S+2\Sigma_{i}(b'^{i}_{0}-k_{0}\delta^{i}_{GS})ln\,[\eta(T_{i})^{2}]
\la{d41}\ee
where $\eta$ is the Dedekind-eta function
($\eta(T)=q^{1/24}\Pi_{n=1}^{\infty}(1-q^{n}), q=e^{-2\pi T}$)
and $b_{0}$
the one-loop beta function for the hidden gauge group. The coefficient
$\delta_{GS}$ is the Green-Schwarz term needed to cancel the gauge
independent part of the target modular anomaly and $b'^{i}_{0}
=\frac{1}{16\pi^{2}}(C(G_{0})-\Sigma_{R_{0}}h_{R_{0}}T(R_{0})(1+2n^{i}_{R_{a}}))$
where $n^{i}$ is the modular weight for a chiral matter superfield
with respect to the $i$-moduli, $C(G_{0})$ the quadratic Casimir operator and
$h_{R_{0}}$ the number of
chiral fields in a representation $R_{0}$ for the hidden sector gauge group.
Through the equation of
motion of  the  auxiliary field of $\Phi$, the scalar component is given in
terms
of the gaugino bilinear  of the hidden sector \ci{r52}
\[
\phi=\frac{e^{-K/2}\xi}{2 \Pi_{i}\eta^{-2}(T_{i})}\,
\bar{\lambda}_{R} \lambda_{L}
\]
with $\xi=2b_{0}/3$.
The model described in eqs.(\ref{d38}-\ref{d40}) is anomaly free and
duality invariant.  The duality transformation for the fields read
\bea
S&\rightarrow& S + 2\Sigma_{i}(k_{0}\delta^{i}_{GS}-b'^{i}_{0})ln(icT_{i}+d),
\nonumber\\
T_{i} &\rightarrow& \frac{a T_{i}-ib}{icT_{i} + d},
\la{d42}\\
\phi&\rightarrow& \phi,
\nonumber\eea
with $a,b,c,d\, \epsilon\, Z $ and $ad-bc=1$.
This model generates
a four-Gaugino interaction and  reproduces the tree
level scalar potential used by other parameterizations of  gaugino
condensate \ci{r50,r51,r55}. It  also  permits the   determination of the
radiative
corrections and use  of NJL technique  \ci{r10} to extract non-perturbative
information in the regime of strong coupling.
After minimizing the complete scalar potential (tree
level plus one-loop potential),  the vacuum structure
is quite different than the tree level one, thus  permitting us to
 find a stable
solution for the dilaton with the inclusion of a single  gaugino
condensate.  The value of the v.e.v.'s of the
dilaton and moduli fields give a good prediction of the fine
structure constant  at the unification scale and unification scale
allowing for minimal string unification  to work.

In supergravity theory the tree level scalar potential is given by \ci{r15}
\be
V_{0}= h_{i}(G^{-1})^{i}_{j}h^{j}-3m^{2}_{3/2}
\la{f8}\ee
where the auxiliary fields are
$h_{i}=-e^{G/2}G_{i}+\frac{1}{4}f_{i}\bar{\lambda}_{R}\lambda_{L}$.
For the choice of Kahler potential and gauge kinetic function given in
eqs(\ref{d38}-\ref{d40}) one has
\be
V_{0}=e^{G}\,B_{0}=
\frac{1}{4}e^{K}\Pi_{i}|\eta(T_{i})|^{-4}\,\,|\phi|^{2}\,B_{0}
\la{d35}\ee
with
\be
B_{0}=
(1+\frac{Y}{\xi})^{2}+\Sigma_{i}\frac{Y}{Y+a_{i}}(1-\frac{a_{i}}{\xi})^{2}\,
\frac{T^{2}_{ri}}{4\pi^{2}}|\hat{G}_{2}(T_{i})|^{2}-3,
\la{d36}\ee
 $a_{i}=2(k_{a}\delta_{GS}^{i}-b'^{i}_{a})$
and  $Y=S+\bar{S} + 2\Sigma_{i}(k_{0}\delta_{GS}^{i}-b'^{i}_{0})lnT_{ri}$
gives the gauge coupling constant at the unification scale. The gravitino mass
is given by
\be
m^{2}_{3/2}=\frac{1}{4}e^{K}\Pi_{i}|\eta(T_{i})|^{-4}\,\,|\phi|^{2}.
\la{f11}\ee
For a fixed value of $S$ the extremum solution to eq.(\ref{d35}) for the moduli
fields gives a
v.e.v. of $<T_{i}>\simeq 1.2$ \ci{r49,r50}.  This value is independent of $S$
and yields an unification scale of order of the string scale much larger than
the required value.
Furthermore, there is
no stable solution in the dilaton
direction, it is a runaway potential for $S\rightarrow \infty$ and it
is unbounded from below for $S\rightarrow 0$. This is not surprising
because we have not included the contribution from loop corrections of
the strong coupling
constant responsible for the gaugino binding. These contributions can be
 calculated using the Coleman-Weinberg one-loop potential
 $V_{1}$  and it is given by  \ci{r23,r57},
\be
V_{1}=\frac{1}{32\pi^{2}} Str \int d^{2} p\, p^{2}
ln(p^{2}+M^{2}) \la{a24}\ee
where $M^{2}$  represents the square mass matrices and $Str$ the
supertrace.
By solving the mass gap equation
$\frac{\partial}{\partial \phi}(V_{0}+V_{1})=0$ one is effectively
summing an infinite number of gaugino bubbles.

The extremum equations for the dilaton and moduli  fields
yield the constraint  \ci{r56}
\be
B_{0}=\frac{9}{2\gamma
b_{0}^{2}}\lr(1+\frac{2\alpha_{0}-1}{3\alpha_{0}-1}\,\epsilon\rr)^{-1}
\la{e30}\ee
and
\be
\epsilon\equiv
x_{g}ln(x_{g})|_{min}=\frac{4b_{0}}{Y}(3\alpha_{0}-1)
\la{e31}\ee
where $B_{0}$ is given in eq.(\ref{d36}),
 $x_{g}=\frac{m^{2}_{g}}{\Lambda_{c}^{2}}$ with  $m^{2}_{g}=m^{2}_{3/2}
\frac{\xi^{2} B_{0}^{2}}{4 (Ref)^{2}}$ the
gaugino mass  square and $\Lambda_{c}$ the condensation
scale given by
\be
\Lambda_{c}=\Lambda_{gut}\, e^{-k_{0}Y/4b_{0}}
\ee
and
\be
\Lambda_{gut}=M_{s}(\Pi_{i}T_{ri}|\eta(T_{ri})|^{4})^{\alpha^{i}_{0}/2}
\ee
 the unification scale.  The coefficient $\alpha_{0}$ is given by
\be
\alpha_{0}^{i}\equiv \frac{\delta^{i}_{GS}k_{0}-b'^{i}_{0}}{b_{0}}
\la{f13}\ee
where the  $i$-index refers to the moduli $T_{i}$.
The extremum solutions eqs.(\ref{e30}) and (\ref{e31})  are obtained
once  $\phi$ has been  integrated  out  ($\phi=\Lambda_{c}^{3}$).  As a
condition of
minimization  for the moduli fields
one gets that the extremum equations are satisfied if  they either take the
dual invariant values ($<T>=1,e^{i\pi/6}$) or they take a common
``large'' value (i.e. $<T_{i}>=<T_{j}>$). For those moduli $T_{i}$ that take a
``large''
v.e.v. the corresponding $\alpha_{0}^{i}$ parameter defined in eq.(\ref{f13})
must be
the same, i.e. $\alpha_{0}^{i}=\alpha_{0}^{j}$.

After eliminating $\phi$, the gravitino mass becomes
\be
m_{3/2}^{2}=\frac{1}{4Y}\,M^{6}_{s}\Pi_{i}(T_{ri}|\eta(T_{i})|^{4})^{3\alpha^{i}_{0}-1}\,e^{-3Y/2b_{0}}
\la{ff3}\ee
and
\be
x_{g}=(\frac{b_{0}}{6})^{2}
\frac{B^{2}_{0}}{Y}\Pi_{i}(T_{ri}|\eta(T_{i})|^{4})^{\alpha_{0}^{i}-1}e^{-Y/2b_{0}}.
\la{f16}\ee
{}From eqs.(\ref{e30}) and (\ref{e31}) it follows that for reasonable solutions
the dominant term
in $B_{0}$ is given by the contribution from v.e.v. of the auxiliary
field  of the dilaton $h_{s}$ and one  can approximate
$B_{0}\simeq (\frac{3Y}{2b_{0}})^{2}$. The v.e.v. of $Y$
and $T_{i}$ are then given in terms of the dimension of the hidden gauge
group, its  one-loop N=1 $\beta$-function coefficient and
$\alpha$ by
\be
Y\simeq
8\pi\sqrt{\frac{1}{n_{g}}\lr(1+\frac{2\alpha_0-1}{3\alpha_0-1}\,\epsilon\rr)^{-1}},
\la{e32}\ee
\be
\Pi_{i}T_{ri}|\eta(T_{i})|^{4}=\frac{\xi^{3}B_{0}^{3}m_{3/2}}{32\,Y\,x^{3/2}_{g}}.
\la{e33}\ee
Eq.(\ref{e33}) is obtained from eqs.(\ref{ff3}) and (\ref{f16})
and the sum
 is over all moduli that acquire a
v.e.v. different from the dual points.  To leading order the v.e.v of the
moduli is
\be
\Sigma_{i}(1-\alpha_{0i})T_{ri} =\frac{3Y}{\pi b_{0}}.
\ee
If the gauge group is broken down from $E_{8}$ to a lower rank group
such as  $SU(N)$ with $5\le N \le 9$,
as can be easily done by compactifying on an orbifold with Wilson lines
 a large hierarchy  can be obtained with  only one gaugino
condensate \ci{r56}.

Having determined the vacuum we now  study  the Yukawa interactions.
The functional dependence of the Yukawa couplings upon the moduli can
be explicitly calculated for specific orbifolds \ci{r60}, but a
generic and useful  way to determine these couplings is to use target space
modular invariance. The Kahler potential given in eq.(\ref{d38})
\be
G=K_{0}+K^{i}_{i}|\varphi_{i}|^{2}+ln\frac{1}{4}|W|^{2}
\la{f33}\ee
 must be duality invariant.
$K_{0}$ is the contribution from the dilaton and moduli fields (cf.
eq.(\ref{d38})) and \ci{r36}
\bea
K^{i}_{i}&=&\Pi_{j}\,d_{j}\,T_{rj}^{n^{ij}},
\la{f34}\\
W&=&W_{0}+W_{m}
\la{f35}\eea
where $d_{j}$ is a constant (which is usually one), $W_{0}$ is given
in eq.(\ref{d39}) and $W_{m}$ is the chiral
superpotential. Modular invariance is achieved if
the chiral superfield $\varphi_{i}$ and superpotential transform as
\bea
\varphi_{i} &\rightarrow& (icT+d)_{j}^{n^{ij}}\varphi_{i},
\la{f36}\\
W  &\rightarrow& \Pi_{i}\frac{W}{(icT+d)_{i}}
\la{f37}\eea
where $n^{ij}$ is the modular weight of the chiral superfield
$\varphi_{i}$ with respect to the $T_{j}$ moduli and the superpotential
transform  as a modular function with weight -1 for each moduli.

For generic abelian $Z_{N}$ and $Z_{N}\,\times\,Z_{M}$ orbifolds, the
modular weights of the chiral superfields vary in  an absolute range
of $-9<\,n\,<4$ depending on which gauge group and Kac-Moody level
they correspond \ci{r34}. Knowing the transformation properties of the chiral
matter fields one can easily derive the transformation property of
the Yukawa couplings $Y_{ijk}$ of the trilinear superpotential
$W_{3}=Y_{ijk}\varphi_{i}\varphi_{j}\varphi_{k}$ and it is given by
\ci{r40,r41}
\be
Y_{ijk} \rightarrow \Pi_{p}(icT+d)_{p}^{-(1+n_{pi}+n_{pj}+n_{pk})}\,Y_{ijk}
\la{f38}\ee
where $n_{pi}$ is the modular weight of $\varphi_{i}$ with respect to
$T_{rp}$.

As an application of this results let us consider a possible
 light scalar fields. Such a field would be phenomenologically important
because if it couples to ordinary  matter, a fifth force mediated by
the exchange of this particle will be generated  which may
 compete with gravity. The
range of the interaction is inversely proportional to the mass of the
exchanged particle. Only if it has a small mass would the range
of this interaction be long enough to be experimentally detected.
A good candidate for a light scalar field is the imaginary part of the
moduli $T_{i}$, $ImT$, whose corresponding real part gets a ``large'' v.e.v.,
because its mass is exponentially suppressed relative to the gravitino
mass. Using the full potential ($V_{0}+ V_1$) the mass for this field
can be calculated (we have taken into account the
fact that $ImT$ does not have  canonical kinetic term) and it is
\be
m_{ImT}=2\pi^{2}\sqrt{\frac{2}{3}}\,T_{r}^{2}\,e^{-\pi T_{r}/2}\,m_{3/2}.
\la{f39}\ee
Clearly, a small mass requires a large v.e.v. for $T_{r}$. Using the
example presented in \ci{r56,r66} for an $SU(6)$
gauge group with $\beta_{0}=15/16\pi^{2}$ which gives a  good prediction of
the gauge coupling constant
at the unification scale ($g_{gut}^{-2}=2.1$) and unification scale ($\simeq
10^{16} \,GeV$),
 the v.e.v. of the moduli fields are
  $T_{r}=44.5,\,24.2,\,17.3$ for
$n_{l}=1,2,3$ respectively ($n_{l}$ counts the number of moduli with
large v.e.v.). The mass in eq.(\ref{f39}) is
$m_{ImT}=2 \times10^{-24},
\,2.4\times10^{-11},\,4.1\times10^{-7}\,GeV$. Therefore a
long range fifth force is only possible for $n_{l}=1$ since for
$n_{l}=2,3$ the range $r\le 8 \times10^{-4}\,cm$ is too small to be
 experimentally detected.

The interaction between $ImT$ and matter fields is given by the Yukawa
interaction
\be
L_{Y}=\frac{1}{2}e^{K/2}\,h_{ijk}\varphi_{i}\bar{\chi}_{j}\chi_{k}
\la{f40}\ee
where $\chi_{i}$ is the fermion component of the $\varphi_{i}$ superfield and
we have used the same symbol for the superfield and its scalar
component. Without going into any specific orbifold example, one can
deduce the functional dependence of the Yukawa couplings on the
moduli, up to a modular invariant function, by simply imposing
duality invariance. Using the duality transformation in eq.(\ref{f38}) one
has
\be
h_{ijk}=\Pi_{p}\,\eta(T_{p})^{-2(1+n_{pi}+n_{pj}+n_{pk})}.
\la{f41}\ee
To compare the strength of the Yukawa interaction relative
 to gravity, we take the two fermion fields in eq.(\ref{f40}) to be quarks
while the scalar one we take as the Higgs doublet. The relative
strength is then given by
\be
\frac{G_{5}}{G_{N}}=\frac{|\tilde{h}_{ijk}<H>|^{2}}{M^{2}_{n}}
\la{f42}\ee
where
$\tilde{h}_{ijk}=\frac{1}{2}e^{K/2}\sqrt{(K^{-1})^{i}_{i}(K^{-1})^{j}_{j}(K^{-1})^{k}_{k}}\,\,h_{ijk}$
is the normalized Yukawa coupling\footnote{$\tilde{Y}$ is given at the
compactification scale and should be run down to the electroweak
scale. We do not worry about this because one expects the change
 to be small}
and $M_{n}$ the nucleon mass. Taking $<H>=300\,GeV$ and $M_{n}=1\,
GeV$, one finds an upper limit for the allowed Yukawa couplings
\be
\tilde{h}_{ijk}\le 10^{-4},
\la{f43}\ee
in order not to contradict the present experimental limit
$G_{5}/G_{N}\le 10^{-3}$ \ci{r62}. Of course eq.(\ref{f43}) only restricts
those Yukawa couplings that depend on the light $ImT$ and in this case
the modular weights of the chiral matter fields are constrained.
Defining $\delta_{i}=T_{ri}|\eta(T_{i})|^{4}$  the
normalized Yukawa coupling can be written as
\be
|\tilde{h}_{ijk}|=\frac{1}{2\sqrt{Y}}\,\Pi_{p}\delta_{p}^{-\frac{1}{2}(1+n_{pi}+n_{pj}+n_{pk})}
\la{f44}\ee
using eq.(\ref{f34}) and setting $d_{i}=1$.
The constraint equation (\ref{f43}) for the
Yukawa couplings using eq.(\ref{e33}) gives
\be
\Sigma_{p}^{n_{l}}(1+n_{pi}+n_{pj}+n_{pk})\le -
2\,\frac{ln(2\times10^{-4}\sqrt{Y})}{ln(\frac{\xi^{3}B^{3}_{0}}{32\,Yx^{3/2}_{g}}\,m_{3/2})}
\la{f45}\ee
where the sum now is over  the moduli that acquire a ``large'' v.e.v.
For phenomenological acceptable solutions the right hand side of eq.(\ref{f45})
is very much restricted. In fact,
the gravitino mass must be of the order of  $10^{2}\, GeV$, $Y\simeq 2$
and  $\xi, B_{0}$ and  $x_{g}$ take a small range of values.
As an example, we choose an $SU(6)$ gauge group in the hidden sector with
$b_{0}=15/16\pi^{2}$ which gives
a  coupling constant at the unification scale of
$g_{gut}^{-2}=2.1$ and allows for an unification scale of the order of $10^{16}
\, GeV$ consistent with MSSM and
$m_{3/2}\simeq 147\,GeV$. Eq.(\ref{f45})   gives
\be
\Sigma_{p}^{n_{l}}(n_{pi}+n_{pj}+n_{pk})\le -2
\la{f46}\ee
for one ``large'' moduli. This last equation
indicates that ordinary matter fields must have negative modular
weights. Considering an overall modular weight defined by
$N_{i}=\Sigma_{p}n_{ip}$ the allowed modular weights of the fields
must satisfy $N_{i}+N_{j}+N_{p}\le -2$. For
abelian orbifolds with standard choice of Kac-Moody levels the allowed
range for an overall modular weights for the standard model fields is
given by $-3\le\,n_{Q,U,E}\,\le 0$ and
$-5\le\,n_{L,D,H,\bar{H}}\,\le1$ \ci{r34}.
It is interesting
to note that the permitted values of the modular weights from a fifth
force consideration given in eq.(\ref{f45}) lie within the range of
 allowed values for the
standard model fields and that this range is further restricted. For any
particular orbifold model, the modular weights of ordinary matter must
satisfy eq.(\ref{f46}) for those Yukawa couplings that depend on
moduli with ``large'' v.e.v.  and this restricts the possible string
vacua.

To conclude, we have shown that  after supersymmetry is broken the v.e.v. of
the
moduli can be  large enough  to produce  a detectable
  fifth force. The absence of such an
interaction imposes non-trivial constraints on any phenomenological acceptable
string vacuum.


{99}

\end{document}